\documentstyle[twocolumn,aps,epsfig]{revtex} 
\begin{document}
\draft
\title{Frustration in $d$-wave Superconducting Circuits: $\pi$-Ring Behaviour}
\author{J.J. Hogan-O'Neill$^{\dagger *}$, J.F. Annett$^{\dagger}$ and 
A.M. Martin$^{\S}$}
\address{$^\dagger$H.H. Wills Physics Laboratory, Tyndall Avenue, Clifton, 
Bristol, BS8 1TL, U.K. }
\address{$^{\S}$D\'{e}partement de Physique Th\'{e}orique, 
Universit\'{e} de Gen\`{e}ve, 1211 Gen\`{e}ve 4, Switzerland}
\date{\today}
\wideabs{
\maketitle
\begin{abstract}
A closed superconducting circuit containing an odd number of $\pi$-junctions,
a $\pi$-ring, has a finite current in the ground state.
We explicitly construct such rings for $d$-wave superconductors and 
demonstrate the existence of spontaneous currents by direct
self-consistent solutions to the Bogoliubov de Gennes equations.
We show that the current has a topological origin due to the 
frustration of the $d$-wave order parameter, which is only partially 
explained by the Sigrist-Rice tunneling formula.
\end{abstract}
} 


It has very recently become clear that
the flux quanta in a superconducting ring are in general of the form $\Phi = (n+\gamma)\Phi_o$,
where $\Phi_o=h/2e$, rather than simply $\Phi = n\Phi_o$ as stated in
the textbooks. The constant $\gamma$ is analogous to the
{\em Maslov index} in Bohr-Sommerfeld quantization 
for single particle dynamics.   The prediction\cite{Geshkenbein,Sigrist}
and subsequent experimental observation\cite{Tsuei1,Kirtley1,Kirtley2,Tsuei2}
 of half-integer flux quantization ($\gamma=\frac{1}{2}$)
has been one of the 
most dramatic developments in the field of superconductivity in recent years.
This discovery is profound for several reasons. 
The experiments provided clear and direct confirmation
of the existence of $d$-wave superconductivity in these materials.
Previously, Wollman {\it et al.}\cite{Wollman} 
had observed a closely related effect: split
Fraunhofer interference peaks in
corner SQUIDs. However, the tri-crystal ring 
experiments\cite{Tsuei1,Kirtley1,Kirtley2,Tsuei2}
were the first to measure non-integral flux quantization directly.
But the implications of non-integral flux quantization 
go beyond simply confirming the pairing state symmetry,
and have not been fully explored.   
The theoretical possibility of rings with flux quanta
other than $\gamma=1/2$ or $\gamma=0$ has been discussed
recently,\cite{Yip} and this has not yet been observed.  
Rings with half-integer flux quantization also have some unique properties,
including an exactly two-fold degenerate ground state. This
opens up new possibilities for novel superconducting devices, which may
find applications in quantum computing\cite{Ioffe,Blais,Schulz} or elsewhere. 
The topology of half-integer quantization has been discussed
recently by Volovik\cite{Volovik}.

The original prediction of the half-integer flux quantization in
$d$-wave superconducting rings was by
Sigrist and Rice{\cite{Sigrist}}, following earlier work 
for the $p$-wave case\cite{Geshkenbein}.
They considered a ring containing a 
single grain boundary junction, and assumed that the
junction current was of the form
\begin{equation}
I = I_{o} \cos(2\theta_{1})\cos(2\theta_{2})\sin{(\varphi)}
\label{eq:sigrist}
\end{equation}
for $d_{x^2-y^2}$ pairing,
where $\varphi$ is the order parameter phase difference across the
grain boundary junction, and  $\theta_{1}$ and $\theta_{2}$ 
are the misalignment
angles between the grain boundary normal and the crystal axes
on either side of the junction.
Provided that these angles are such that the junction is a
$\pi$-junction 
(i.e. the pre-factor of $\sin(\varphi)$ is negative in Eq.~\ref{eq:sigrist})
and the ring inductance is large enough
then the ring will exhibit a spontaneous current
in the ground state.

However, Eq.~\ref{eq:sigrist} probably does not apply.
The grain boundary junctions 
have a complicated micro-structure and 
measured critical 
currents
are approximately exponential in the misalignment angles\cite{Ivanov}.
Also the current-phase relationship of the junctions
is not simply given by $\sin{(\varphi)}$\cite{Ilichev}.
Furthermore the concept of a single $\pi$-junction is not well
defined since gauge transformations make $\varphi$ ill defined
for a single junction\cite{Sigrist}. 
Gauge invariant definitions 
can only be made for closed rings, and so  
one should talk about 
$\pi$-rings  rather than $\pi$-junctions.  

In this letter we show that the existence of $\pi$-rings
stems directly from the frustration of the $d$-wave order parameter
in certain ring topologies.   The frustration 
can only be resolved by a spontaneous 
ground state super-current.  Our calculations are the first
to show this phenomenon at a 
microscopic quantum mechanical level.  We show that for
a frustrated $d$-wave  ring there
is a meta-stable energy maximum with zero current which has
a zero energy state (ZES) at the Fermi level.  In the ground state
the ZES is removed, but at the cost of
a spontaneous circulating current. 
In contrast non-frustrated $d$-wave rings 
have no ground state current.
Our numerical results for grain boundary junctions 
also differ considerably from the
Sigrist-Rice model
Eq.~\ref{eq:sigrist};
the current is not proportional to $\sin{\varphi}$,
and the angle dependence is not 
given by $\cos{(2\theta_1)}\cos{(2\theta_2)}$.

First let us review the symmetry principles
underlying the Josephson characteristics of grain boundary junctions
in $d_{x^2-y^2}$ superconductors. Assuming time reversal
symmetry, the current can be expanded\cite{Annett_Ger}
\begin{equation}
  I(\theta_1,\theta_2,\varphi) = \sum_n I_n(\theta_1,\theta_2) \sin{(n\varphi)}.
  \label{eq:sineseries}
\end{equation}
For tetragonal lattice symmetry the functions $I_n(\theta_1,\theta_2)$
are obviously periodic in $\theta_1$ and $\theta_2$
with period $\pi$.
 Rotation of one crystal by $\pi/2$ changes the sign of the 
$d$-wave order parameter, and so we have a symmetry
$\theta_1 \rightarrow \theta_1+\pi/2$, $\varphi \rightarrow \varphi+\pi$. 
The implications of this symmetry are different for current harmonics with
$n$ even or odd,
\begin{figure}
\centerline{\epsfig{figure=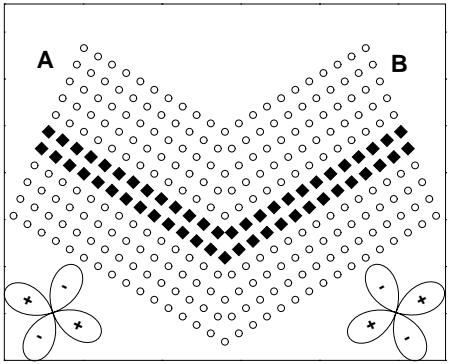,width=6cm}}
\caption{a) The large angle grain boundary ring. The order parameter configuration 
shown is for $0^o$ phase difference across the grain boundary, b) The $[110]$ ring. 
The orientation of the order parameter lobes is for a phase difference of $0^o$ 
across the $[110]$ boundary.}
\label{pos32}
\end{figure}
\begin{eqnarray}
  I_n(\theta_1+\frac{\pi}{2},\theta_2) & = & - I_n(\theta_1,\theta_2) \hspace*{1cm}
   {\rm n \ odd} \label{eq:nodd} \\
     I_n(\theta_1+\frac{\pi}{2},\theta_2) &= &+ I_n(\theta_1,\theta_2) 
     \hspace*{1cm}  {\rm n \ even.} \label{eq:neven}
\end{eqnarray}
The odd $n$ harmonics have sign changes at certain angles,
but the even $n$ harmonics need not. 
The existence of $\pi$-junctions is therefore not guaranteed by
symmetry alone.
Our earlier calculations\cite{Hogan}
and experiments of Il'ichev {\it et al.}\cite{Ilichev} show that
the $n > 1$ harmonics cannot be ignored.
These higher harmonics give a
$4e/h$ Josephson coupling at angles where the $2e/h$ coupling
vanishes{\cite{Tanaka94}}.
In the weak coupling 
Josephson regime where only the
$n=1$ current component is significant the  critical current 
$I_c = I_1(\theta_1,\theta_2)$
and this can be expanded as a Fourier series{\cite{Walker}}
 \begin{equation}
I_c =C\cos{2\theta_{1}}\cos{2\theta_{2}} +
S\sin{2\theta_{1}}\sin{2\theta_{2}} + \dots . \label{eq:series}
\end{equation}
The Sigrist-Rice form, Eq.~\ref{eq:sigrist}, 
is merely the first expansion term.  If the other terms are significant
then  one cannot predict whether a given junction orientation
has $\pi$-junction or normal junction behaviour
merely from the sign of $\cos{2\theta_1}\cos{2\theta_2}$.
Also, in contrast to Eq.~\ref{eq:sigrist}, the current
does not vanish for the case $\theta_1=\theta_2 = \pi/4$, corresponding to
a twin boundary in a slightly orthorhombic crystal.

In our calculations we directly compute the 
current in $d$-wave superconducting circuits
by solving the Hartree-Fock Gorkov or Bogoliubov-de Gennes (BdG) 
equations {\cite{Martin97}.
We use a tight binding lattice with nearest neighbor hopping, $t$,
and retarded attractive interaction $U_{ij}=-3.5t$
with a  frequency cutoff  $E_{c}=3.0t$.
We used a temperature of $T=0.01t$ and a chemical potential of $\mu=0$.
The calculation{\cite{Martin97} is self-consistent in 
the non-local order-parameter, 
$\Delta_{ij}=U_{ij}\langle c_{i \uparrow}c_{j\downarrow}\rangle$,
and in the hopping, $t_{ij} =  t + \frac{1}{2} U_{ij} n_{ij}$, where
$n_{ij}=\sum_{\sigma}\langle c_{i \sigma}^{\dagger}c_{j \sigma}\rangle$.
Currents are computed 
for each nearest neighbor bond\cite{Hogan}, $I_{ij}$, and self-consistency
ensures that current conservation is obeyed at each site.
\begin{figure}
\centerline{\epsfig{figure=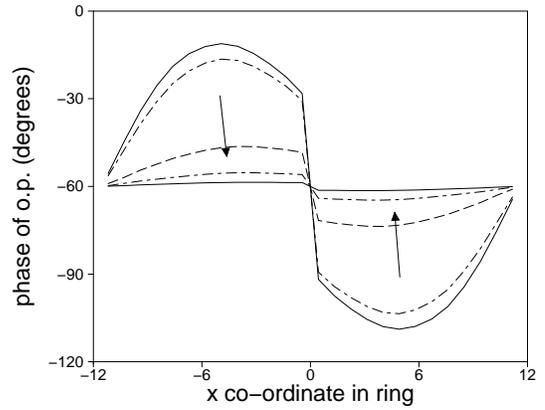,width=7cm}}
\caption{Evolution of order parameter phase within the ring. The arrow direction 
indicates increasing self-consistent iteration number so that by the time 
self-consistency is achieved we almost have a perfectly flat line (no phase 
gradient) around the sample.}
\label{faze32}
\end{figure}

To address the question of the origin of $\pi$-ring
 behaviour at a microscopic quantum mechanical 
level we have directly calculated the current in rings
of $d$-wave superconductor.  An example is shown in Fig.~\ref{pos32}.
This model grain boundary junction corresponds to misorientation
angles of $\theta_1=\theta_2 = \tan^{-1}{1/2} = 26.6\deg$.
In earlier work we have studied the critical current  and current-phase
characteristics of such junctions in bulk superconductors\cite{Hogan}. Here we
have constructed superconducting rings containing these junctions.
This is done by simply connecting the edges labeled
{\bf A} and {\bf B} in Fig.~\ref{pos32}.  The top and bottom edges 
shown in the figure cannot be connected, and so they are left 
with open boundary conditions. This provides a tractable
microscopic model corresponding to the tri-crystal ring experiments
of Tsuei and Kirtley\cite{Tsuei1,Kirtley1,Kirtley2,Tsuei2}.
Of course the model has only one grain boundary, not three, but that
is sufficient to demonstrate the principle.  The other two
junctions in the tri-crystal can be viewed simply as a means of connecting
the two 
faces {\bf A} and {\bf B}. 
If the two additional junctions were identical in symmetry
then
their own characteristics would be irrelevant\cite{Walker} and the ring 
characteristics depend only on the single central boundary.
In this ring geometry we have a system 
containing typically 200 or more inequivalent lattice sites
and self-consistency in $n_{ij}$ and $\Delta_{ij}$
must be achieved independently on every bond.

Fig.~\ref{faze32} shows the evolution of the phase of the
$d$-wave order parameter\cite{Martin97,Hogan} in the ring
system of Fig.~\ref{pos32} as 
the system approaches self-consistency.
The phases plotted correspond to the sites indicated in black 
in Fig.~\ref{pos32}. 
Similar results are obtained for the other lines
around the ring.  The starting point for the calculation
was (chosen arbitrarily) a step-function 
phase difference of $\varphi=-120^o$ applied across the grain boundary,
with an opposite step at the {\bf A}-{\bf B} interface.
Fig.~\ref{faze32} shows that
as the iterative self-consistent procedure is carried out 
there is an `averaging-out' across the boundary, i.e the phase on
the right hand side drops and that on the right hand side 
\begin{figure}
\centerline{\epsfig{figure=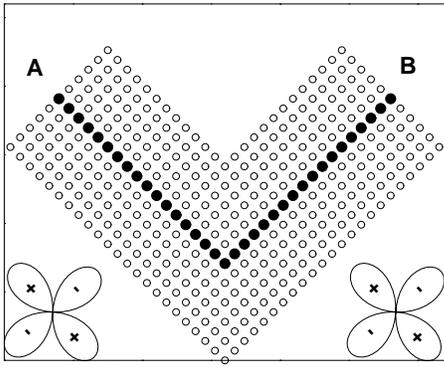,width=6cm}}
\caption{The $[110]$ ring. The orientation of the order parameter lobes is for 
a phase difference of $0^o$ across the $[110]$ boundary.}
\label{pos22}
\end{figure}
increases until
it reaches their mean value of $-60^0$ (a straight line) at
self-consistency. Thus, for this system the ground state
has no ground-state circulating
current, and is therefore not a $\pi$-ring.

The conclusion that this first geometry is not a $\pi$-ring is
not surprising. 
In our earlier
work\cite{Hogan} we calculated the supercurrent-phase characteristic,
$I(\theta_1,\theta_2,\varphi)$ for a grain boundary 
junction between two bulk superconductors for this geometry.
Choosing a gauge
so that zero phase difference corresponds to the 
$d$-wave orientation indicated
in Fig.~\ref{pos32} leads to  a positive current with
a positive phase difference, $\varphi$, as expected for a normal 
junction\cite{Hogan}. 
The sign of the $d$-wave order parameter can be chosen consistently
around the circuit as indicated,  implying
the ring geometry of Fig.~\ref{pos32} has no intrinsic frustration


Now contrast this with the ring geometry shown in Fiq.~\ref{pos22},
where again edges {\bf A} and {\bf B} are joined.
In this geometry the topology clearly indicates that the $d$-wave
order parameter is frustrated\cite{Annett_Ger}. 
Suppose we define the gauge so that zero phase difference across the
junction corresponds to that indicated in Fiq.~\ref{pos22}.
Then one can see that one must introduce a sign change at the 
point where the {\bf A }and {\bf B} edges are joined to make a close ring.
Alternatively in a different gauge convention one could impose no sign change
at the {\bf A}-{\bf B} interface, but then necessarily there must be a sign 
change elsewhere in the loop. This intrinsic frustration is a new 
feature of
the system Fiq.~\ref{pos22}, and is quite different from
Fig.~\ref{pos32} where there was no similar frustration.
Topologically Fiq.~\ref{pos22} is a {\em globally non-orientable surface},
equivalent to the M\"obius strip (one cannot continuously
assign a unique  orientation (gauge) at each point), while
Fiq.~\ref{pos32} is globally 
orientable and topologically
equivalent to the surface of an ordinary cylinder.
See Ref.~\cite{Volovik} for further implications of this topology.

The dramatic effect of this topological difference is apparent
in the evolution to self-consistency, as seen in Fig.~\ref{faze110}.
Starting the evolution to self-consistency
with $\Delta_{ij}=0$ for all sites on the 
left hand side of the boundary and a bulk $d$-wave state with
an arbitrary phase of $30^o$ on the right, we obtain the order parameter phase
\begin{figure}
\centerline{\epsfig{figure=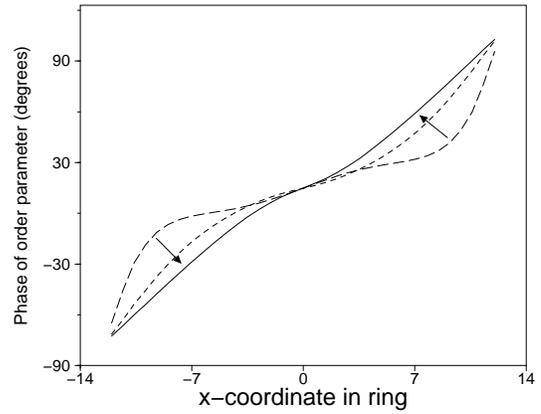,width=7cm}}
\caption{Evolution of the phase difference across the $[110]$ interface with 
increasing self-consistent iteration number. The arrows point in the direction 
of increasing iteration number.}
\label{faze110}
\end{figure}
shown in Fig.~\ref{faze110}.  The phase
becomes increasingly linear, corresponding to a uniform
and non-zero supercurrent throughout the ring. The system has
therefore a {\em spontaneous ground state current}
and is therefore a $\pi$-ring.  Note that the discontinuity in phase
at the join of the {\bf A} and {\bf B} edges is purely due to the
sign convention chosen in Fig.~\ref{pos22}, and does not correspond to any physical 
discontinuity in the $\Delta_{ij}$.
Also the slight
non-linearity in the phase gradient at 
$x=0$ is consistent with the increased cross-section
width of the strip at this point. Fig.~\ref{faze110}
is a direct microscopic example of a $\pi$-ring
due to topological frustration in $d$-wave superconductors.
In order to understand the microscopic origin of this spontaneous current
it is helpful to
first consider a single $\theta_1=\theta_2=45\deg$ junction between two
bulk superconductors.  This would be a $[110]$ twin boundary for two 
slightly 
orthorhombic superconductors.
The geometry is identical to
Fig.~\ref{pos22}  except that the edges {\bf A} and {\bf B} are not joined, but rather
connect to infinite bulk superconductor on either side of the junction.
Now there is translational symmetry parallel to the interface, and every
site is symmetrically equivalent to one of those indicated in black
on Fig.~\ref{pos22}.  In this case we can use the numerical 
methods of Ref.~\cite{Hogan} to determine the characteristics of this junction.
We choose to define zero phase difference $\varphi$ as indicated in 
Fig.~\ref{pos22}.
In this case the non-local order parameter $\Delta_{ij}$ clearly has
a perfect bulk $d$-wave shape at every site, including the 
sites at the center of the boundary.
Calculating the local density of states for this 
does indeed yield an ideal bulk $d$-wave density of states.

Suppose now that we apply a phase difference $\varphi$ across the 
interface.   Rotating the order parameter on the right hand side of the 
interface by $90^o$  corresponds to a phase difference of $\varphi=180^o$.
In this case the local density of states
is shown in Fig.~{\ref{block}}(a). 
There is a strong zero energy state (ZES), or resonance, at the center of the
gap.  This ZES is essentially of the same origin as the one
described by  Belzig, Bruder and Sigrist\cite{Belzig}. 
\begin{figure}
\centerline{\epsfig{figure=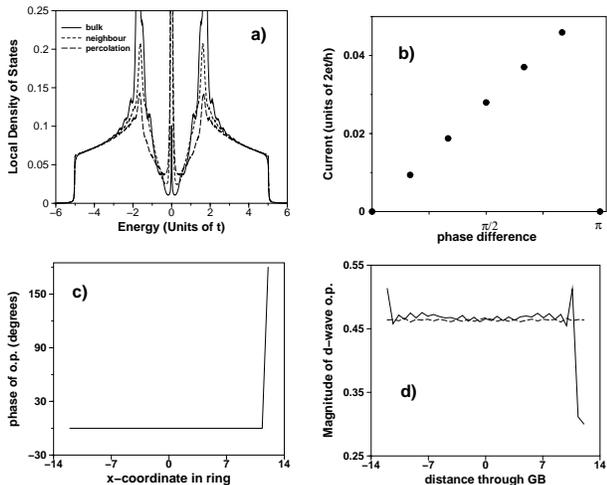,width=8cm}}
\caption{a) LDOS for the $[110]$ interface in bulk with a phase difference of $180^o$, 
b)Current versus phase for same boundary, c) $0^o$ phase difference across the 
$[110]$ interface when joined into a ring, d) $|\Delta_{d}|$ for same system as c): 
full line is for $30^o$, broken line for $0^o$.}
\label{block}
\end{figure}
However in our case
we do not observe splitting of the peak, and there is no
time reversal symmetry breaking (TRSB). 
The final self-consistent answer yields a $d$-wave order parameter but 
with a small extended-$s$ component at this phase difference. The pairing state at the junction is locally $s+d$ and not $s+id$. 
We have also calculated the supercurrent-phase characteristic of this
$[110]$ twin boundary junction. We 
apply a phase difference $\varphi$
between the two bulk superconductors and calculating self-consistently the 
order parameter $\Delta_{ij}$, charge $n_{ij}$ and 
current $I_{ij}$ on each bond in the region of the junction. 
Fig.{\ref{block}}(b) shows that the current versus phase profile 
is an almost linear saw-tooth function 
with sharp discontinuities at $\pm \pi$. It is qualitatively the same as 
for the other boundaries previously reported{\cite{Hogan}}. 
The shape is indicative of a strong coupling junction{\cite{Likh}}, and 
we attribute the sharp discontinuity
to the existence of the ZES at $\varphi=\pi$.
Obviously the fact that the curve is not simply a sine wave indicates that
the higher harmonics are relevant in Eq.\ref{eq:sineseries},
consistent with the experiments of Il'ichev 
{\it et al.}\cite{Ilichev}.
Also note that in this junction geometry the Sigrist-Rice formula 
Eq.~\ref{eq:sigrist} predicts no supercurrent flow at all, but 
the non-zero current is due to the $\sin{2\theta_1}\sin{2\theta_2}$ or
higher order terms in the expansion Eq.~\ref{eq:series}.

The state $\varphi=\pi$ in Fig.~\ref{block}(a) with the ZES 
is the energy maximum for the junction, as can be seen by
integrating the current, $\int I d\varphi$, of Fig~\ref{block}(b).
It corresponds to a {\em phase slip} of $\pi$ in the order parameter
across the junction.  Such states are also possible in the
ring geometry.   Fig.~\ref{block}(c) shows a self-consistent
order parameter phase for the same ring geometry of Fig.~\ref{faze110}.
This was obtained by starting the self-consistent calculation
with a bulk $d$-wave order parameter and with the phase arrangement 
shown in Fig.~{\ref{pos22}}
Thus, the calculation begins with a {\it phase-slip}
at the {\bf A}-{\bf B} boundary. At this point the magnitude of the
$d$-wave order parameter $|\Delta_d|$ is heavily suppressed, as shown in
Fig.~\ref{block}(d). 
This is because locally there are sites where the $\Delta_{ij}$
to the four neighboring sites are $(+,-,-,+)$ rather than
$(+,-,+,-)$ counting clockwise. 
We interpret this phase slip solution for the ring as the meta-stable 
energy maximum
state which separates the two equivalent energy minima
corresponding to positive or negative circulating current.


In summary we identify ring geometries in $d$-wave 
superconductors as either frustrated or not. 
Frustrated  rings have a circulating current in the ground state 
for  but no ZES.  The only available states without current flow
have a phase slip, but with the onset of a 
non-split ZES (i.e. no time reversal symmetry breaking). We also calculated 
the current 
versus phase profile and demonstrated that the Sigrist Rice formula is not a 
valid approximation. Finally, we suggest that experiments 
designed to measure the differential conductance in these rings are an 
important next step in understanding the role of the ZES.

We would like to thank J.P. Wallington and B.L. Gy{\"{o}}rffy for useful
discussions. This work was supported by 
the EPSRC under grant GR/L22454
and the TMR network Dynamics of Nanostructures.

* Present address: Laboratory of Physics, Helsinki University of Technology, 
P.O. Box 1100, FIN-02015 HUT, Finland.

\end{document}